\documentclass{PoS}

\title{The $N_f= 2$ chiral phase transition from imaginary chemical potential with Wilson Fermions}

\ShortTitle{The $N_f= 2$ chiral phase transition from imaginary chemical potential}

\author{\speaker{Christopher Pinke}\\
        Institut für Theoretische Physik - Johann Wolfgang Goethe-Universität\\
        Max-von-Laue-Str. 1, 60438 Frankfurt am Main, Germany\\
        E-mail: \email{pinke@th.physik.uni-frankfurt.de} }
\author{Owe Philipsen\\
        John von Neumann Institute for Computing (NIC)\\ GSI, Planckstr. 1,
        64291 Darmstadt, Germany\\ 
        Institut für Theoretische Physik - Johann Wolfgang Goethe-Universität\\
        Max-von-Laue-Str. 1, 60438 Frankfurt am Main, Germany\\
        E-mail: \email{philipsen@th.physik.uni-frankfurt.de} }

\abstract{
The order of the thermal transition in the chiral limit of QCD with two dynamical flavours of quarks is a long-standing issue. 
Still, it is not definitely known whether the transition is of first or second order in the continuum limit. 
Which of the two scenarios is realized has important implications for the QCD phase diagram and the existence of a critical endpoint at finite densities. 
Settling this issue by simulating at successively decreased pion mass was not conclusive yet. 
Recently, an alternative approach was proposed, extrapolating the first order
phase transition found at imaginary chemical potential 
to zero chemical potential with known exponents, which are
induced by the Roberge-Weiss symmetry. 
For staggered fermions on $N_t=4$ lattices, this results in
a first order transition in the chiral limit. 
Here we report of $N_t=4$ simulations with Wilson fermions, 
where the first order region is found to be large.
}

\FullConference{The 33rd International Symposium on Lattice Field Theory\\
                 14 -18 July  2015\\
                 Kobe International Conference Center, Kobe, Japan}

\usepackage[utf8]{inputenc}
\usepackage[T1]{fontenc}
\usepackage{amsmath, amssymb}
\usepackage[format=hang, font=small,labelfont=bf,textfont=it, justification=centerlast, margin=1mm]{caption}

\newcommand{\ii}{\ensuremath{\textrm{i}}}
\newcommand{\Tr}{\operatorname{Tr}}

\newcommand{\NGroup}{\ensuremath{\text{N}}}
\newcommand{\NSigma}{\ensuremath{\text{N}_\sigma}}
\newcommand{\NTau}{\ensuremath{\text{N}_\tau}}
\newcommand{\Nc}{\ensuremath{\text{\NGroup}_\text{c}}}
\newcommand{\Nf}{\ensuremath{\text{\NGroup}_\text{f}}}
\newcommand{\VSpatial}{\ensuremath{V}}

\newcommand{\Loewe}{LOEWE-CSC}
\newcommand{\Lcsc}{L-CSC}
\newcommand{\clqcd}{CL\kern-.25em\textsuperscript{2}QCD}
\newcommand{\codename}{\clqcd}

\newcommand{\psibar}{\bar{\psi}} 
\newcommand{\chiralcond}{\ensuremath{\langle \psibar \psi \rangle}}
\newcommand{\mpi}{\ensuremath{m_{\pi}}}
\newcommand{\mpiC}{\ensuremath{\mpi^c}}
\newcommand{\MFermion}{\ensuremath{D}}

\newcommand{\MFermionMinus}{\ensuremath{\MFermion^{-1}}}

\newcommand{\Order}{\ensuremath{\mathcal{O}}}

\newcommand{\LatSpacing}{\ensuremath{a}}
\newcommand{\LatMass}{\ensuremath{m}}
\newcommand{\LatCoupling}{\ensuremath{\beta}}
\newcommand{\LatCouplingC}{\ensuremath{\LatCoupling_c}}

\newcommand{\Binder}{\ensuremath{B_4}}
\newcommand{\Skew}{\ensuremath{S}}
\newcommand{\Temp}{\ensuremath{T}}
\newcommand{\Tc}{\ensuremath{\Temp_c}}
\newcommand{\Mu}{\ensuremath{\mu}}
\newcommand{\MuI}{\ensuremath{\mu_I}}
\newcommand{\MuIc}{\ensuremath{\MuI^c}}
\newcommand{\GenObs}{\ensuremath{X}}

\begin{document}

\section{Introduction}
\label{ch:introduction}

	\begin{figure}
		\centering
		\begin{minipage}{.45\linewidth}
					\includegraphics[width=\linewidth]{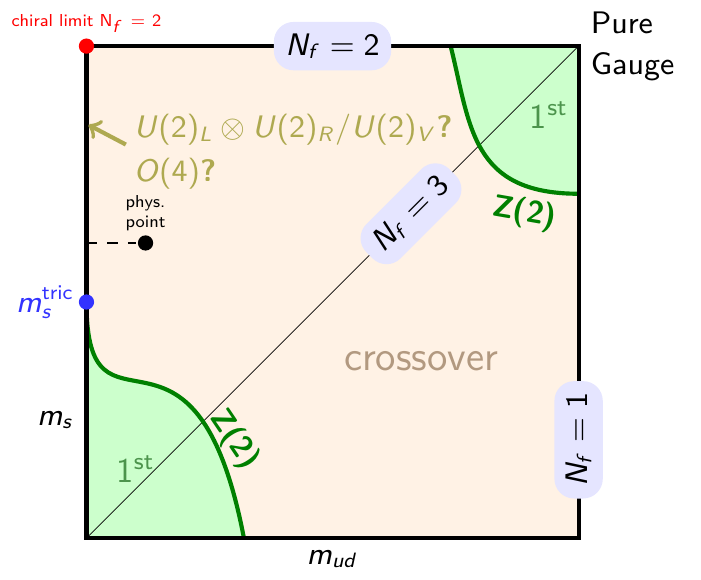}
		\end{minipage}
		~~~
		\begin{minipage}{.45\linewidth}
					\includegraphics[width=\linewidth]{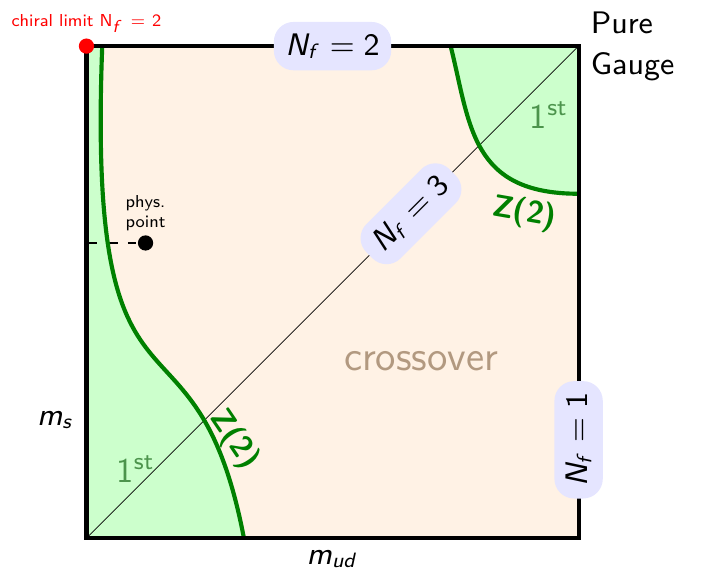}
		\end{minipage}
		\caption{Possible scenarios for the QCD phase diagram at $\Mu=0$ as function of quark mass.}
		\label{fig:columbiaPlot}
	\end{figure}

The clarification of the order of the thermal transition in the chiral limit of $\Nf=2$ QCD is a long-standing issue. 
Still, it is not definitely known if the transition is of first or second order. 
This is depicted in Figure \ref{fig:columbiaPlot}.
Which of the two scenarios is realized has important implications for the physical QCD phase diagram, and in particular it is important regarding the possible existence of a critical endpoint at finite densities. 
Settling this issue by simulating at successively decreased pion mass was not conclusive yet, primarily because of the increasing demands of the simulations as the pion mass is lowered (see e.g. \cite{Bonati:2014kpa} for references and a more detailed introduction to the topic).

\begin{figure}
		\centering
		\begin{minipage}{.45\linewidth}
					\includegraphics[scale=.7]{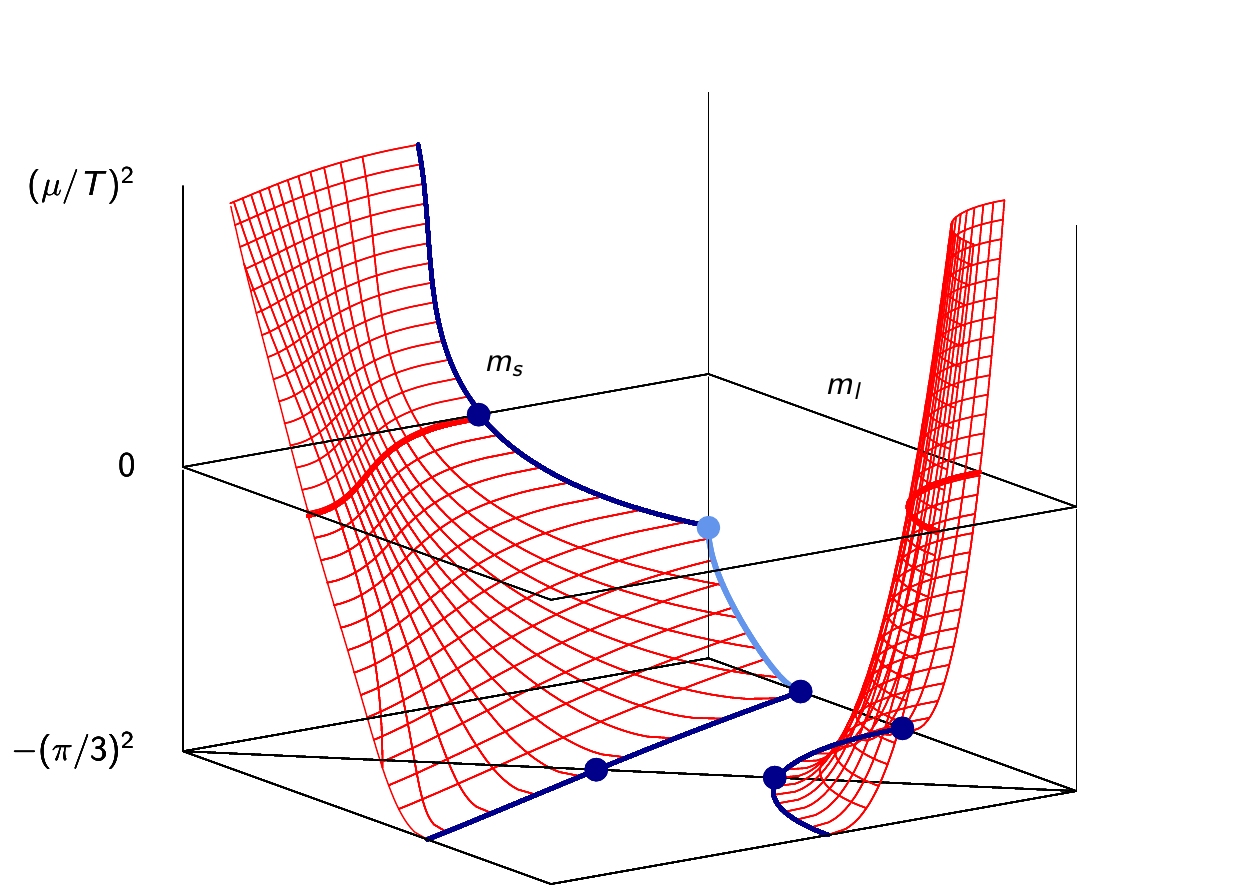}
		\end{minipage}
		~~~
		\begin{minipage}{.45\linewidth}
					\includegraphics[scale=1]{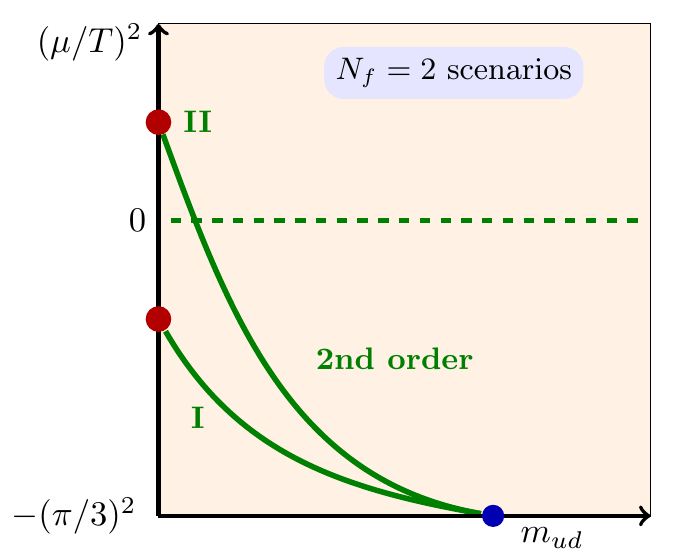}
		\end{minipage}
		\caption{Left: Expected QCD phase diagram as function of $(\Mu/T)^2$. Right: Possible scenarios in the $\Nf=2$ backplane. Both figures follow \cite{Bonati:2014kpa}.}
		\label{fig:columbiaPlot3dAndNf2Backplane}
	\end{figure}

Recently, an alternative approach was proposed \cite{Bonati:2014kpa}, which relies on the nontrivial phase structure of QCD at purely imaginary chemical potential $\mu_I$.
In this region of phase space, one has reflection symmetry and extended center (or Roberge-Weiss (RW)) symmetry \cite{Roberge:1986mm}:
\begin{eqnarray}
	\label{ZPeriodicityMui}
	Z(\mu)&=&Z(-\mu),\\ 
Z \left(\Mu \right)&=&Z \left (\Mu + 2\pi\ii k/\Nc\ \right),\ k \in \mathbb N\;. 
\label{RWSymmetry}
\end{eqnarray} 
At imaginary chemical potential $\mu=i\mu_I$ the sign problem is absent and standard simulation algorithms can be applied.
Critical values of $\MuIc = (2k+1)\pi/\Nc\ (k \in \mathbb N)$ mark the boundary between adjacent center sectors.
The transition between these sectors in $\mu_I$ is first order for high and 
crossover for low temperatures.
The endpoint of this so-called Roberge-Weiss transition 
meets 
with the chiral/deconfinement transition continued from real $\Mu$.
Thus it corresponds to 
a triple point at low and high masses, where first order chiral and
deconfinement transitions join it, and to a second order endpoint 
for intermediate masses, where the thermal transition is just a crossover. 
These regions are separated by tricritical points. 
When mass and $\Nf$ are changed at fixed $\MuIc$, a 
phase diagram similar to that shown in 
Figure \ref{fig:columbiaPlot} for $\mu=0$ emerges.
Both are analytically connected when $\mu_I$ is varied, which is depicted in 
Figure \ref{fig:columbiaPlot3dAndNf2Backplane} (left).
More specifically, leaving the critical $\MuI$-values (the bottom of the figure), lines of second order transitions depart from the tricritical points, separating regions of first order transitions from crossover regions. 
In the vicinity of the tricritical points, the line is governed by tricritical scaling laws, which allows for an extrapolation to the chiral limit. 
For the $\Nf=2$ backplane, which is of interest in this study, the possible scenarios are shown in Figure \ref{fig:columbiaPlot3dAndNf2Backplane} (right).
If the tricritical point at $m=0$ is at negative values of $\Mu^2$, the chiral phase transition is second order.
On the other hand, if it is at positive values, there exists a first order region at $\Mu=0$ and the transition in the chiral limit must be first order, too.
In this way one can clarify the order of the chiral limit at zero chemical potential by mapping out the second order line.
Using staggered fermions on $N_t=4$ lattices, it was indeed found that the transition is of first order in the chiral limit \cite{Bonati:2014kpa}. 
These findings have to be contrasted with other fermion discretizations. 
We report on the status of our simulations with Wilson fermions following the same approach.

\section{Simulation details}
\label{ch:simulationDetails}

We employ the same numerical setup as for the study of the Roberge-Weiss
transition described in \cite{Philipsen:2014rpa},
with the standard Wilson gauge action
and two flavours of mass-degenerate unimproved Wilson fermions.
The bare fermion mass \LatMass\ is encapsulated in the hopping parameter $\kappa = (2(\LatSpacing \LatMass + 4))^{-1}$.
Finite temperature on the lattice is given by the inverse temporal extent, $\Temp = 1/\left( \LatSpacing(\LatCoupling)\NTau \right)$.
All simulations were carried out using the OpenCL\footnote{See www.khronos.org/opencl for more information.} based code \codename\ \cite{Bach:2012iw}\footnote{Which is now available at {github.com/CL2QCD}\;.}, which runs efficiently on Graphic Processing Units (GPUs) on \Loewe\ at Goethe-University Frankfurt \cite{Bach2011a} and on \Lcsc\ at GSI Darmstadt \cite{lcsc}.
We work at fixed temporal lattice extent $\NTau = 4$, leaving the RW-plane $\MuIc=\pi\Temp$ investigated in \cite{Philipsen:2014rpa}.
In the latter study, the lowest mass simulated was at $\kappa = 0.165$. 
Here we add $\kappa = 0.170, 0.175$ and $0.180$.
In order to locate the critical chemical potential for each bare quark mass, simulations at various values of the quark chemical potential $\LatSpacing\Mu$ were carried out.
For each of these parameter sets, temperature scans were carried out with $\Delta\LatCoupling$ at least $0.001$ around the critical coupling on three spatial volumes, $\NSigma=12, 16$ and 20, corresponding to aspect ratios $\NSigma/\NTau$ of 3,4 and 5, respectively.
After discarding 5k to 10k trajectories for thermalization, 40k to 60k trajectories have been simulated on each individual Monte-Carlo chain, such that there are at least 100 independent measurements around the critical region.
The autocorrelation on the data was estimated using a python implementation\footnote{See github.com/dhesse/py-uwerr .} of the Wolff method \cite{Wolff:2003sm}.
To accumulate statistics faster, we simulated four chains for each parameter set.
Observables are measured after each trajectory and the acceptance rate in each run was of the order of $75\%$.
Additional \LatCoupling-points have been filled in using Ferrenberg-Swendsen reweighting \cite{Ferrenberg:1989ui}.

\section{Results}
\label{ch:results}

	\begin{figure}
		\centering
		\begin{minipage}{.45\linewidth}
					\includegraphics[scale=0.6]{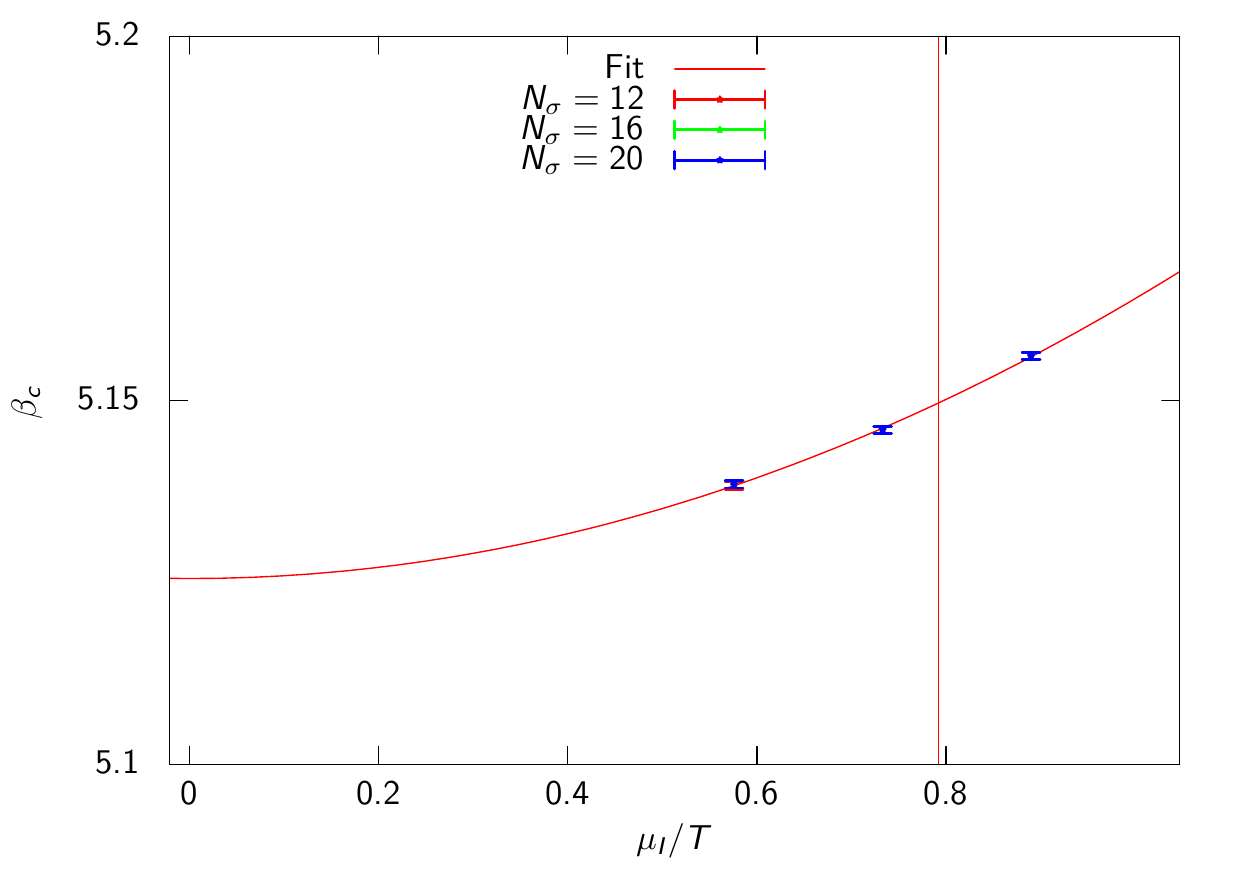}
		\end{minipage}
		~~~
		\begin{minipage}{.45\linewidth}
					\includegraphics[scale=0.6]{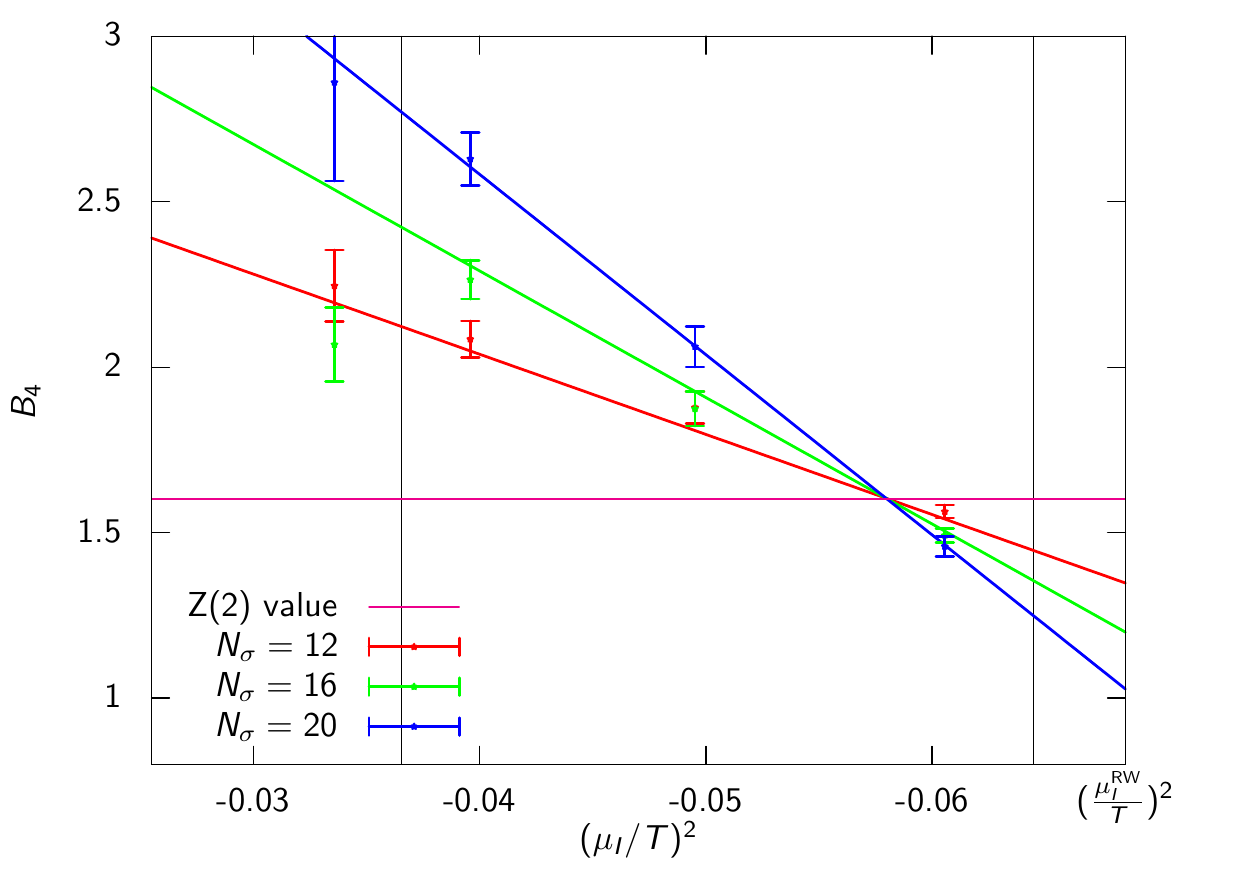}
		\end{minipage}
		\caption{Left: Critical coupling as function of $\Mu/\Temp$ for $\kappa=0.170$. The line indicates $\LatCouplingC(\MuIc)$. Right: Finite size scaling of \Binder\ and fit for $\kappa=0.165$. The vertical lines indicate the fit ranges.}
		\label{fig:results1}
	\end{figure}

	\begin{figure}
		\centering
		\begin{minipage}{.45\linewidth}
					\includegraphics[scale=0.6]{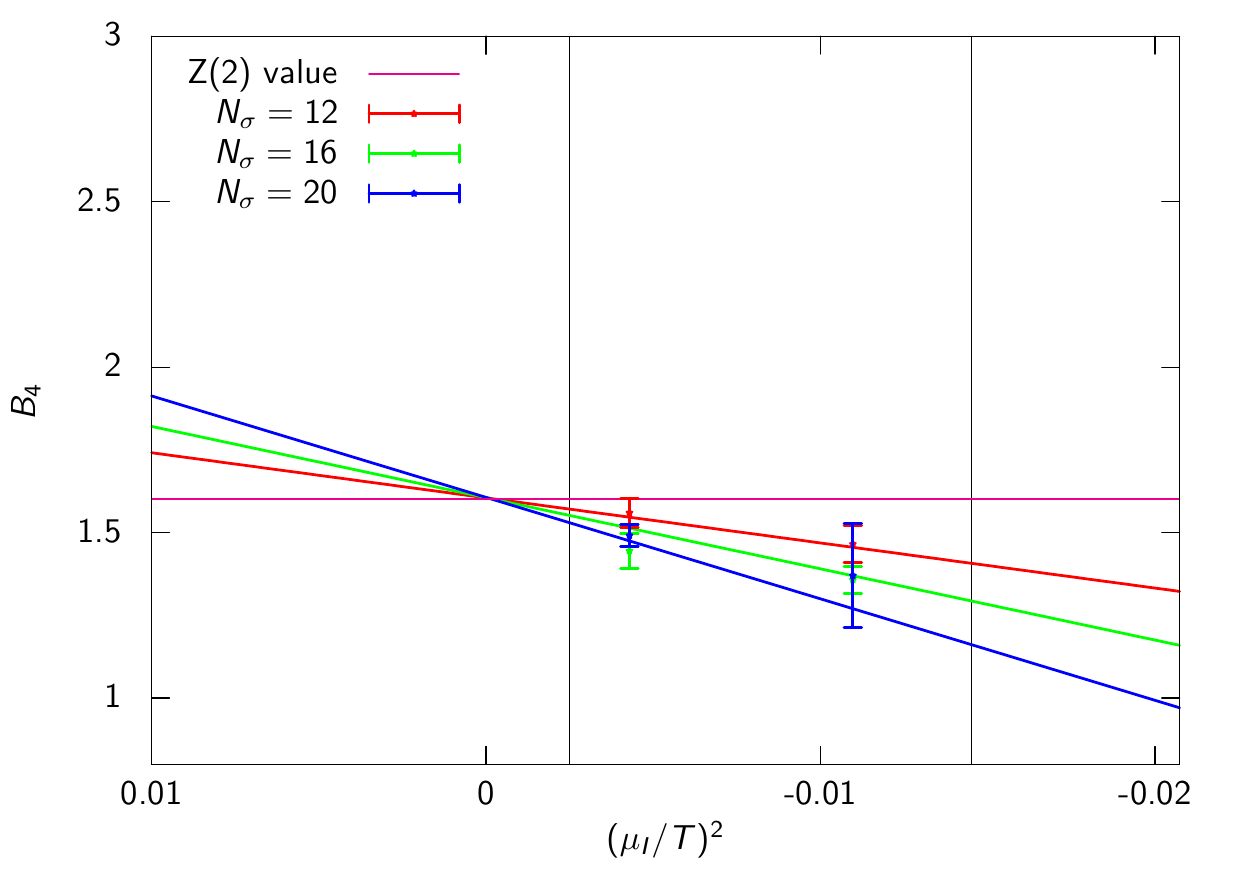}
		\end{minipage}
		~~~
		\begin{minipage}{.45\linewidth}
					\includegraphics[scale=0.6]{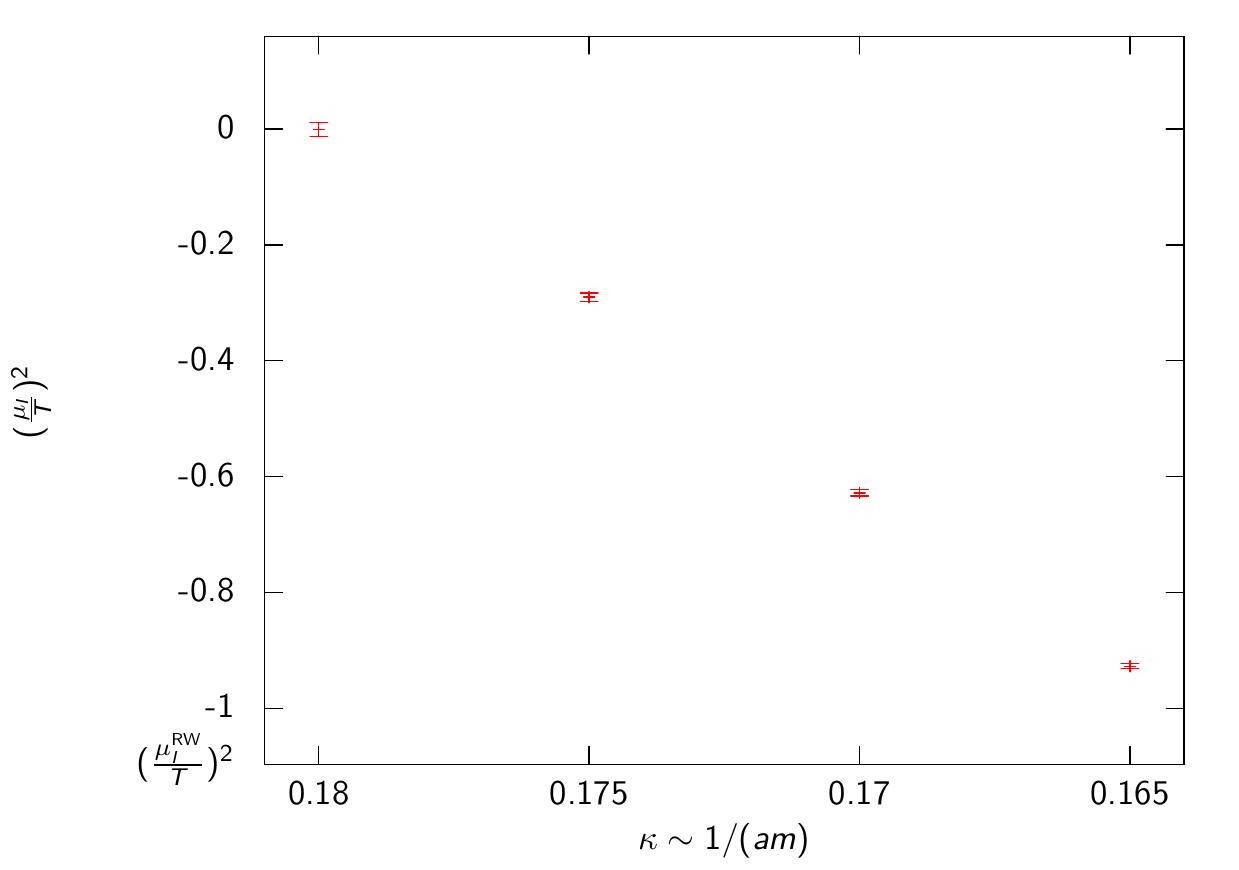}
		\end{minipage}
		\caption{Left: Preliminary fit results for $\kappa=0.180$. Right: Results for $\MuIc(\kappa)$.}
		\label{fig:results2}
	\end{figure}

We define a (pseudo-)critical temperature \Tc\ or coupling \LatCouplingC\ by the vanishing of the skewness
\begin{equation}
	\Skew\ =  \langle (\GenObs - \langle \GenObs \rangle)^{3}\rangle /  \langle	(\GenObs - \langle \GenObs \rangle )^{2}	\rangle^{3/2}
\end{equation}
of a suitable observable \GenObs.
In this study, we use the chiral condensate $\chiralcond = \Nf\ \Tr{\MFermionMinus}$.
Figure \ref{fig:results1} (left) shows \LatCouplingC\ for $\kappa = 0.170$.
For a fixed $\kappa$, the $\LatCouplingC(\Mu)$ can be nicely fitted to a quadratic, even function, as shown in the figure.
This allows to interpolate between simulation points and to extrapolate towards zero chemical potential.
As expected, the results show a decreasing critical temperature as the chemical potential approaches zero.
The same holds if the (bare) mass is lowered.

In the thermodynamic limit $\VSpatial \rightarrow \infty$, the Binder cumulant \cite{Binder:1981sa}
\begin{equation}
	\Binder(\GenObs) = \langle (\GenObs - \langle \GenObs \rangle)^{4}\rangle /  \langle	(\GenObs - \langle \GenObs \rangle )^{2}	\rangle^{2}
	\label{bindercum}
\end{equation}
allows to extract the order of the transition.
In particular, it takes the values 1 for a first order transition and 3 when there is no true phase transition but a crossover.
For the case of a second order transtion in the $3D$ Ising universality class, it has a value of around $1.604$ (see e.g. \cite{blote}).
Hence, a discontinuity exists when passing from the first order to the crossover region via the second order endpoint.
To clarify the type of transition on finite lattices one can look at the finite-size scaling of the \Binder\ at \LatCouplingC. 
In the vicinity of the second order point it scales as \cite{Binder:1981sa}
\begin{align}
	\Binder(\LatCouplingC, \NSigma) &= b_1 + b_2 \left[(\LatSpacing\MuI)^2 - (\LatSpacing\MuIc)^2\right] \NSigma^{1/\nu} + \ldots \;.
	\label{BinderScaling}
\end{align}
For the 3D Ising universality class, one has $b_1 \approx 1.604$ and $\nu \approx 0.63$.
The critical coupling \MuIc\ indicates the position of the $Z(2)$ point.

The values for $\Binder(\LatCouplingC, \NSigma)$ obtained from the simulations can be fitted to this form.
Examples are given in Figure \ref{fig:results1} (right) and \ref{fig:results2} (left).
The results for $\MuIc$ are shown in Figure \ref{fig:results2} (right).
Results for \LatCouplingC\ and \MuIc\ are summarized in Table \ref{tab:critBetaAndMu}.
As can be seen in the figures, \Binder\ increases with volume if the transition is a crossover (left of the second order point), whereas it decreases in the first order region, ultimately approaching the infinite volume values 3 and 1, respectively.
As the (bare) quark mass is lowered, $\MuIc/\Temp$ decreases towards zero (see Figure \ref{fig:results2} (right)).
Note that the results for $\kappa=0.180$ are preliminary.
With the given data, one has to extrapolate to \MuIc\ (see Figure \ref{fig:results2} (left)).
Therefore we are currently adding more simulations here in order to get a better estimate of \MuIc.
Nevertheless, \MuIc\ is clearly compatible with zero within errors given the current data.
This indicates that also for $\Mu=0$ a first order region is present.

\begin{table}
\captionsetup{width=5cm}
\centering
\makebox[0pt][c]{\parbox{.65\textwidth}{%
		\hspace*{-2.5cm}
    \begin{minipage}[l]{0.32\hsize}\centering
					\begin{tabular}{|c|c|c|}
					\hline
					$\kappa$ & \LatCouplingC & $\LatSpacing\MuIc$ \\ 
					\hline
					0.165 & 5.2421(1) & 0.2408(6) \\ \hline
					0.170 & 5.1497(9) & 0.1981(9) \\ \hline
					0.175 & 5.0519(3) & 0.1346(17) \\ \hline
					0.180 & 4.9520(2) & 0.0090(423) \\ \hline
					\end{tabular}
        \caption{Results for \LatCouplingC\ and $\LatSpacing\MuIc$ from finite size scaling.}
        \label{tab:critBetaAndMu}
    \end{minipage}
		\hspace*{2.5cm}
\captionsetup{width=8cm}
    \begin{minipage}[r]{0.32\hsize}\centering
					\begin{tabular}{|c|c|c|c|c|}
					\hline
					$\kappa$ & $\LatCoupling$ & a[fm] & T [MeV] & $\mpi$ [MeV] \\ \hline
					0.1800 & 4.9519 & 0.309(3) & 162(2) & 587(6) \\ \hline
					0.1750 & 5.0519 & 0.301(3) & 166(2) & 642(7) \\ \hline
					0.1700 & 5.1500 & 0.288(3) & 174(2) & 699(7) \\ \hline
					0.1650 & 5.2420 & 0.271(3) & 185(2) & 770(8) \\ \hline
					0.1575 & 5.3550 & 0.246(3) & 203(2) & 929(10) \\ \hline
					\end{tabular}
        \caption{Results of the scale setting. See text for details.}
        \label{tab:scaleSetting}
    \end{minipage}
}}
\end{table}

To relate these findings to physical scales we performed a series of $\Temp=0$ simulations at or close to the respective \LatCouplingC\ for each $\kappa$.
Similar to \cite{Philipsen:2014rpa} we generated $\Order(400)$ independent configurations on $32^3\times12$ lattices.
With these at hand, we set the scale by the Wilson flow parameter 
$w_0$ using the publicly available code described in \cite{Borsanyi:2012zs}.
This method is very efficient and fast.
In particular, it is much more precise than setting the scale by the $\rho$ mass, which was done in \cite{Philipsen:2014rpa}.
Hence, we revaluated the $\Temp=0$ simulations from this study, carried out at $\kappa=0.1575$, and include them here for completeness.
In addition, the pion mass \mpi\ was determined\footnote{We thank F. Depta for carrying out these measurements.}.
The results for the lattice spacing \LatSpacing, the critical temperature \Tc\ and \mpi\ in physical units are summarized in Table \ref{tab:scaleSetting}.
The results show that, in terms of pion masses, the first order region is large.
Note that the lattices coarsen going to lower masses, since \LatCouplingC\ decreases.
However, all lattices considered are very coarse, $\LatSpacing \gtrsim 0.25$ fm.
Because of this, huge discretization artifacts can be expected.
Note that $\mpi L > 5$ holds for all our parameter sets, so that finite size effects are negligible.

\section{Discussion \& Perspectives}
\label{ch:discussion}

\begin{figure}
 		\centering
   \includegraphics[scale=0.7]{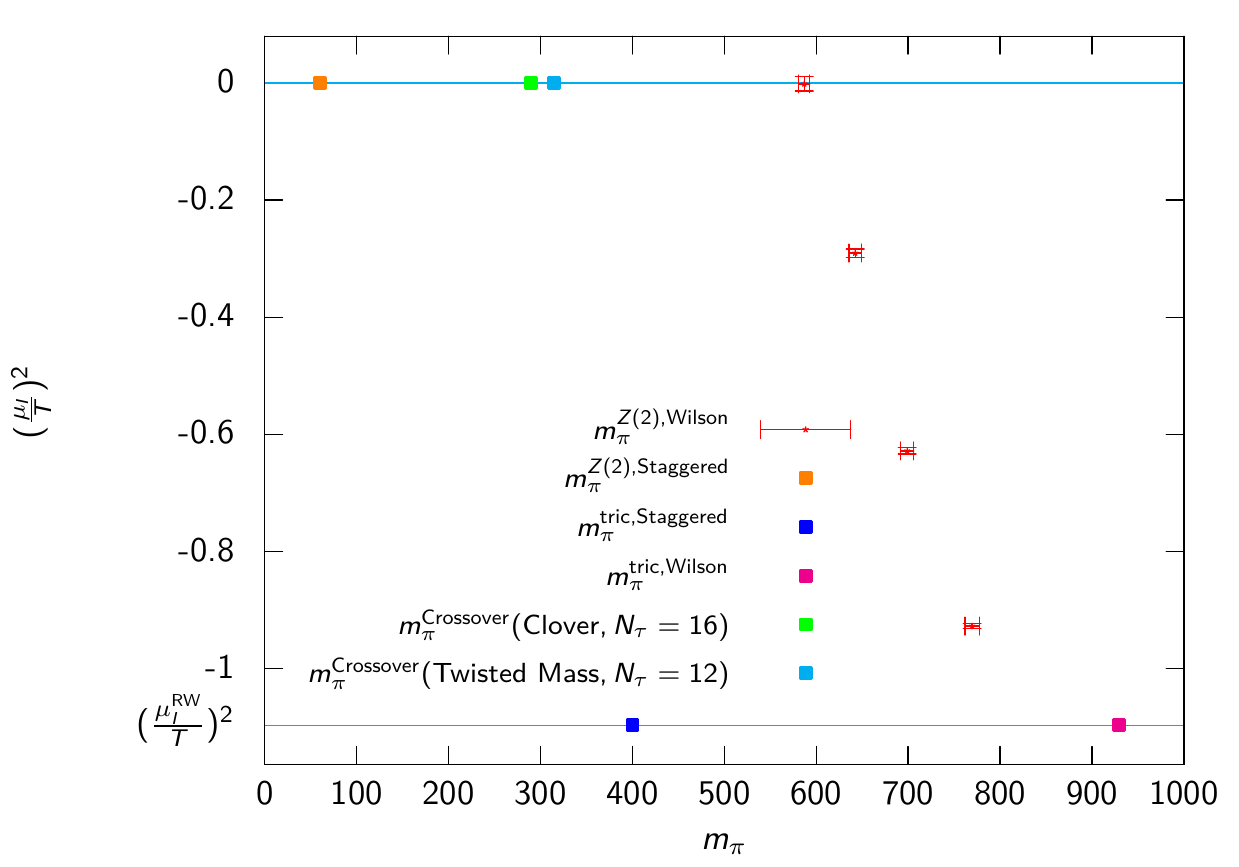}
 		\caption{$Z(2)$ line in the \mpi-$(\Mu/T)^2$ plane. See text for details.}
 		\label{fig:mpiVsMuSq}
\end{figure}
Our findings are summarized in Figure \ref{fig:mpiVsMuSq}, which shows the $Z(2)$ line in the \mpi-$\Mu^2$ plane.
Leaving the RW-plane, the critical line approaches zero $\Mu$ at clearly non-vanishing pion masses.
In fact, the lowering of the critical pion masses is steady and relatively mild.
This finding corresponds to the second scenario sketched in Figure \ref{fig:columbiaPlot3dAndNf2Backplane} (right).

For comparison, the results from the previous study using staggered fermions \cite{Bonati:2014kpa} are also shown in Figure \ref{fig:mpiVsMuSq}.
On coarse lattices, both staggered and Wilson discretizations show similar behaviour with a first order region at $\Mu=0$.
This region is much larger for Wilson fermions.
In ($\Order(\LatSpacing)$-improved) Wilson studies at $\NTau = 12$ \cite{Burger:2011zc} and $16$ \cite{Brandt:2014sna}, also shown in the figure, only crossover signals are seen, yielding an upper bound for a possible first order region.
This suggest that the observed wide first order region is to a large extent due to discretization effects.

To put our $\NTau=4$ results into perspective, they differ from
early results with Wilson fermions based on a different simulation strategy
\cite{iwasaki}. However, they are in accord with modern investigations.
A recent study with $\Order(\LatSpacing)$-improved Wilson-Clover fermions determined a similarly large \mpiC\ of around 880 MeV for $\Nf=3$ on $\NTau=4$ lattices\cite{Jin:2014hea}.
In another study presented at this conference \cite{czaban}, it was shown that the tricritical point in the RW-plane moves to lower masses by only 
100 MeV within our action. 
Taken together, this may suggest that the 
$\Order(\LatSpacing)$ effects are not yet dominant, 
at least not on $\NTau=4$ lattices.
To reduce cut-off effects, one has to study the \NTau-dependence of the first order region.
We will determine this dependence for the $\Mu=0$ endpoint of the chiral critical line in future studies.

\newpage
\acknowledgments

O. P. and C. P. are supported by the Helmholtz International Center for FAIR 
within the LOEWE program of the State of Hesse.
We thank \Loewe\ and \Lcsc\ at GU-Frankfurt and the NIC in J\"ulich
for computer time and support.

\end{document}